\documentclass[twoside,twocolumn]{article}

\usepackage{wscg}           
\RequirePackage{ifpdf}
\ifpdf
 \RequirePackage[pdftex]{graphicx}
 \RequirePackage[pdftex]{color}
\else
 \RequirePackage[dvips,draft]{graphicx}
 \RequirePackage[dvips]{color}
\fi
\usepackage{nopageno}       

\usepackage{censor} 
\usepackage{bm} 
\usepackage{physics} 
\usepackage{amssymb} 
\usepackage{mathtools}	
\usepackage{tabularx} 
\usepackage{algorithm} 
\usepackage[noend]{algpseudocode} 
\usepackage{multirow} 
\usepackage{tabularx,booktabs} 
\usepackage{todonotes}

\usepackage{xcolor}

\newcolumntype{Y}{>{\raggedleft\arraybackslash}X}

\StopCensoring		

\title{\textsc{Leaven} - Lightweight Surface and Volume Mesh Sampling Application for Particle-based Simulations}

\author{
	\parbox{0.35\textwidth}{\centering
		\censor{Alexander Sommer$^{1}$}\\[1mm]
		\censor{alexander.sommer@hs-rm.de}
	}
	\hspace{0.05\textwidth}
	\parbox{0.35\textwidth}{\centering
		\censor{Ulrich Schwanecke$^{1}$}\\[1mm]
		\censor{ulrich.schwanecke@hs-rm.de}
	}
	\\[8mm]
	\parbox{\textwidth}{\centering 
		\censor{$^1$ Computer Vision and Mixed Reality Group, RheinMain University of Applied Sciences} \censor{Wiesbaden R\"usselsheim, Germany}} \\
}

\usepackage{url}
\makeatletter
\def\Uslash{\mathbin{\mathchar`\/}\@ifnextchar{/}{\kern-.15em}{}}
\g@addto@macro\UrlSpecials{\do \/ {\Uslash}}
\def\Ucolon{\mathbin{\mathchar`:}\@ifnextchar{/}{\kern-.1em}{}}
\g@addto@macro\UrlSpecials{\do : {\Ucolon}}
\makeatother

\begin{document}

\twocolumn[{\csname @twocolumnfalse\endcsname

\maketitle  

\begin{abstract}
\noindent
We present an easy-to-use and lightweight surface and volume mesh sampling standalone application tailored for the needs of particle-based simulation. We describe the surface and volume sampling algorithms used in \textsc{Leaven} in a beginner-friendly fashion. Furthermore, we describe a novel method of generating random volume samples that satisfy blue noise criteria by modifying a surface sampling algorithm. We aim to lower one entry barrier for starting with particle-based simulations while still pose a benefit to advanced users. The goal is to provide a useful tool to the community and lowering the need for heavyweight third-party applications, especially for starters.

\vspace{0.5em}


\vspace*{1.0\baselineskip}

\end{abstract}
}]


\section{Introduction}

\copyrightspace{80-903100-7-9}{2005}{January 31 -- February 4}

In many fields of computer graphics, the plausible simulation of different physical phenomena is an important topic. As simulations hit interactive framerates, the urge for a unified simulation solver allowing to simulate various phenomena in one framework grew. A lot of methods like position-based dynamics \cite{PBD07} and their successors unified particle physics \cite{UPP14} and projective dynamics \cite{ProjectiveDynamics2014}\cite{Sommer2020} use particles to represent all various simulation objects, like rigid-bodies, cloth, fluid, granular material, gases, and deformable solids. Further, in other non-real-time simulation methods, like Smoothed Particle Hydrodynamics (SPH) \cite{SPH09} or material point methods \cite{MPM18}, particle representations play an important role. They are not only used for characterizing parts of a continuum but also in the efficient and correct handling of object- and domain boundaries \cite{Akinci2012}.

We experienced that especially for beginners generating a particle representation of different geometrical objects is an entry barrier for getting started with particle-based simulations. There are a certain number of well-described algorithms for generating surface representations, as well as some algorithms for generating suitable volume representations (see Section \ref{sec:relWork}).

Beginners often struggle to find a convenient and ready-to-use implementation or tool that is suitable for the requirements of particle-based simulation. The use of heavy-weight third-party software for sampling can also introduce new issues. These programs can be hard to access and learning how to work around the vast amount of functions can be unnecessarily time-consuming. Furthermore, data-export is often limited to undocumented binary data formats.

Therefore, we introduce \textsc{Leaven}, a Lightweight surfac\textsc{e} \textsc{a}nd \textsc{v}olume m\textsc{e}sh sampli\textsc{n}g application tailored to the needs of particle-based simulations.   \textsc{Leaven} is an easy-to-use graphical application that enables beginners to sample a surface or volume of a 3D triangle mesh with only a few straightforward clicks. Furthermore, it leaves enough customization options to be valuable for experienced simulators. \textsc{Leaven} is made available as open-source and can be used as a library in other projects.

Our paper is organized as follows: first, we give a brief overview of sampling algorithms (Section \ref{sec:relWork}), then we describe the algorithms used for sampling in our application (Section \ref{sec:alg}), followed by a detailed description of the developed application (Section \ref{sec:app}) and its usage (Section \ref{sec:usage}). Our main contributions are:
\begin{itemize}
	\item Giving a beginner-friendly introduction to particle sampling techniques
	\item Lowering the entry barrier for particle-based simulations
	\item Modifying a surface sampling method to generate volume samples that satisfy blue noise criteria
	\item Contributing an easy-to-use, lightweight, open-source standalone application for sampling the surface and volume of arbitrary closed triangle meshes for use in particle-based simulations
\end{itemize}
%
%

\begin{figure*}
	\centering
	\includegraphics[width=1.0\linewidth]{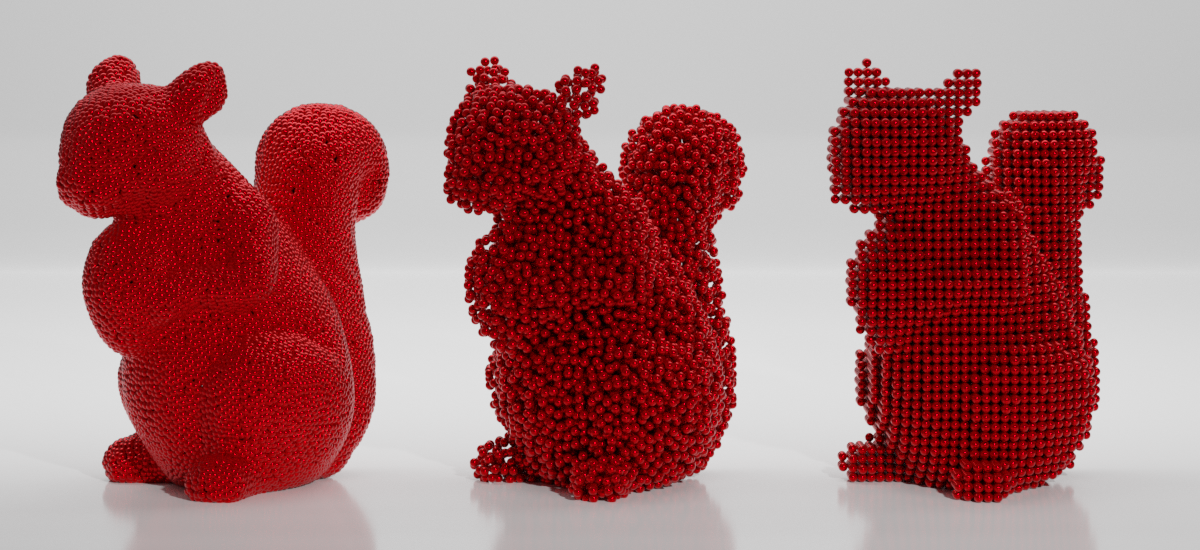}
	\caption{Different samplings in comparison. Surface sampling (left), volume sampling with randomly distributed particles (middle), and volume sampling with grid-based packing (right).}
	\label{fg:samples}
\end{figure*}

\section{Related Work}
\label{sec:relWork}

Various fields of computer graphics, such as texturing, remeshing, rendering, modeling, or animating fractures, have methods for sampling surfaces or volumes that are tailored to their specific needs. An important sampling criterion that many of these methods have in common is that the sampling behaves like blue noise. Blue noise distributions have been an important research topic \cite{Cook1986}\cite{Ostromouknov2004}\cite{Jiating2013}. A sampling that has blue noise characteristics is uniform and unbiased distributed in the spatial domain. Furthermore, when looking at the frequency domain with the Fourier transformed spatial information, there should be a lack of low-frequency noise and structural bias. One popular method for generating samplings that satisfy blue noise criteria is Poisson disk sampling \cite{Mccool1992}\cite{Bridson2007}. In Poisson disk sampling, the samples are placed randomly on the valid spatial domain, which is the surface of an object for surface sampling or its volume for volume sampling. After placing the samples, a validation process guarantees that no two samples are too close to each other. Cline et. al. \cite{Cline2009} introduced an optimized dart-throwing algorithm for generating Poisson disk point sets as a sampling of the surface of a 3D triangle mesh. Their algorithm needs sequential processing and has its applications in modeling and polygon remeshing. Wei et. al. \cite{Wei08} present a parallel dart-throwing algorithm by drawing suitable samples from a dense point cloud on the object's surface. This parallel version can achieve interactive frame-rates even for a large number of samples and is therefore well suited for real-time applications. Bowers et. al. \cite{Bowers10} extended this algorithm to work not only with the Euclidean but also with a fast approximation of the Geodesic norm, as a measurement of sample point distances. This method is well suited for interactive texturing applications or other tasks where quick sampling is needed. Yuksel \cite{Yuksel2015} describes an approach of generating surface samplings without using a form of dart-throwing but rather assigning weights to possible samples. He uses a greedy algorithm to eliminate the points with the highest weight. Wang, T. et. al. \cite{Wang2018} propose another rapid parallel surface sampling method. They use blue-noise pattern planes that they progressively project on the mesh surface using ray tracing. This method has its applications especially when fast sampling is required. A novel data-driven approach by Wang, Z. et. al. \cite{Wang2020} promises to achieve even better performance than the classic geometrical approach on triangle meshes.

In contrast to the previously mentioned techniques, Jiang et. al. \cite{Jiang2015} use different methods inspired by SPH simulations. They are not only generating surface samplings but also volume samplings in that way. Their idea is to treat each sampling point as an SPH particle and establish a constant density in SPH kernel estimation functions by applying correction forces. Adams et. al. \cite{Adams2007} describe an adaptive volume sampling method for particle-based fluid simulation. They use a sampling condition to put more focus on geometrically complex simulation regions. 

\begin{table*}
	\centering 
	\begin{tabularx}{\textwidth}{|l|r| *{6}{Y|}}
		\hline
		\multirow{2}{*}{Model} & \multicolumn{1}{c|}{BBox} &\multicolumn{2}{c|}{Surface Sampling} & \multicolumn{2}{c|}{Random Volume Sampling} & \multicolumn{2}{c|}{Grid Volume Sampling} \\
		& \multicolumn{1}{c|}{Scaling} & \multicolumn{1}{c|}{time} & \multicolumn{1}{c|}{particles} & \multicolumn{1}{c|}{time} & \multicolumn{1}{c|}{particles} & \multicolumn{1}{c|}{time} & \multicolumn{1}{c|}{particles} \\
		\hline
		Bunny & 2.0 & 13ms & 9122 & 293ms & 14875 & 100ms & 22206 \\
		Squirrel & 3.0 & 30ms & 14889 & 573ms & 31795 & 109ms & 48012 \\
		Armadillo & 6.0 & 87ms & 53921 & 5096ms & 145403 & 2559ms & 215141 \\
		Dragon & 12.0 & 595ms & 233355 & 23444ms & 943703 & 6710ms & 1445523 \\
		
		\hline
		
	\end{tabularx}
	\caption{Timing results for different meshes and varying particle numbers resulting from different scales of the bounding box and a fixed particle radius of 0.02m for volume sampling / minimum distance for surface sampling.}
	\label{tbl:timing}
\end{table*}

\section{Algorithms}
\label{sec:alg}

The main algorithms used in \textsc{Leaven} can be divided into surface- and volume sampling algorithms. In particle-based simulations, surface representations are typically used for boundary handling and sometimes for two-way coupling, while volume samplings are used for initial particle positions or rigid-body representations in unified solvers. They both have in common that they need a discrete 3D representation as an input. 3D geometry in computer graphics is often represented as polyhedrons, with a polygon mesh defining the surface of the object. We use triangle meshes as input since these meshes are most common, easy to work with, and accessible. For volume sampling, the input mesh describing the surface of an object must be closed to have a distinguishable inside and outside. Figure \ref{fg:samples} shows different samplings of the same 3D object. On the left the surface sampling algorithm is applied, where particles can get as close as half the particle diameter. Shown in the middle is our approach to generating random blue noise samples within a volume, with a minimum particle spacing equal to the particle diameter. A grid-based volume sampling is shown on the right. Table \ref{tbl:timing} shows timing results for the three algorithms available in \textsc{Leaven} on different standard meshes.

\subsection{Surface Sampling}
\label{subsec:surface}

Akinci et. al. \cite{Akinci2012} describe how boundary particles can be a suitable boundary representation in SPH fluid simulation and how to calculate a two-way fluid-rigid coupling. Later this practice has been used for boundary handling in position-based methods like position-based fluids \cite{PBF2013}, or for the simulation of gases and granular material \cite{Ihmsen2012}. 

There are some basic requirements for a good surface representation with particles. First of all, the sampling should be dense enough that no simulation particle can tunnel through holes between boundary particles. Furthermore, the particles should be spread out uniformly on the object's surface to guarantee an even force distribution to prevent artifacts.

To generate such a sampling of the surface of an object, we use a uniform sampling algorithm on arbitrary triangle meshes introduced by Bowers et. al. \cite{Bowers10}. The method builds upon a grid cell sampling approach to generate Poisson disk samples on arbitrary manifold surfaces \cite{Wei08}. The goal of the sampling process is to find a set of sample points (particles) $\mathbb{S}$ that is randomly but uniformly distributed on the surface $\partial V$ of a 3D Volume $V$. Thereby all particles should have a minimum distance $d$ to each other, i.e.
\begin{equation*}
	\text{dist}(\bm{s}_i, \bm{s}_j) \geq d, \;\; \forall \bm{s}_i, \bm{s}_j \in \mathbb{S}
\end{equation*}

For most cases in particle-based simulation, $d$ equals the radius $r$ of the simulation particles. The distance metric $\text{dist}(\cdot, \cdot)$ measuring the spatial separation can be Euclidean or Geodesic. The Euclidean distance measures the shortest line segment between two points in 3D space. The Geodesic distance measures spatial separation with respect to the object's surface as a sub-manifold. This can be beneficial for achieving a good sampling of complex objects with thin features. For a detailed explanation of Geodesic distances and their fast approximation, we refer the reader to \cite{Bowers10}.

The algorithm starts initially by calculating a much larger set $\mathbb{P}$ with candidate positions on the surface $\partial V$ without bothering about the distances between them. The candidate samples are found by picking a random triangle from the triangle mesh with a probability proportional to the triangle's area. On this triangle, an arbitrary position $\bm{p}$ is generated by choosing two random values $\tau_1, \tau_2 \in [0,1]$ for its barycentric coordinates
$$
	u = 1 - \sqrt{\tau_1},\quad
	v = \tau_2 \sqrt{\tau_1},\quad 
	w = 1 - u - v
$$
inside the triangle, which leads to a random position 
\begin{equation*}
\bm{p} = u \bm{a} + v \bm{b} + w \bm{c},
\end{equation*}
where $\bm{a},\bm{b},\bm{c}$ are the three vertices of the triangle. In this manner, the initial set $\mathbb{P}$ is filled with a large number of candidate positions. In the application \textsc{Leaven}, we generate $\rho_S \cdot \frac{A(\partial V)}{\pi r^2}$ positions in this way, where $A(\partial V)$ is the surface area, which is the sum of all triangle areas and $\rho_S$ is a controllable density parameter. We use a default value of $\rho_S = 40$. After this, the bounding box containing $V$ is sliced into grid cells with a uniform side length of $\frac{d}{\sqrt{3}}$. This leads to a cell diagonal of $d$. For each position, $\bm{p} \in \mathbb{P}$ the corresponding grid cell id is calculated and the whole data array containing $\mathbb{P}$ is sorted by the grid cell id. Each cell that contains at least one position $\bm{p}$ is a valid surface cell. Cells without candidate positions are much likely located outside or inside the volume.

Since the majority of cells is empty a hash table is a suitable lookup structure for finding valid cells and their positions. The cell id is used as the hash key. In contrast to \cite{Bowers10}, we use a spatial hashing function with bit-wise \texttt{xor}:
$$
(i \cdot p_1 \; \texttt{xor} \; j \cdot p_2 \; \texttt{xor} \; k \cdot p_3) \; \% \; \texttt{size(hash\_table)}
$$
where
$$
p_1 = 73856093, \; p_2 = 19349663, \; p_3 = 83492791
$$
as described in \cite{SpatialHashing2003}, where $i,j,k$ are the cell indices, to prevent hash collisions.
Each hash bucket contains the cell id, a pointer to the first position $\bm{p}$ that is associated with this cell id in the concurrent sorted data array, and if already chosen, the sampled position. In the end, each valid surface cell will contain at most one sample.

\begin{algorithm}
	\centering
	\begin{algorithmic}[1]
		\State generateCandidateSetP()
		\State sortSetPByCellId(setP)
		\State insertSetPInHashMap(setP)
		\For {trial $t = 1,\ldots, n$}
		\ForAll {valid cells $i$}
		\If {cell $i$ is already sampled}
		\State \textbf{break}
		\EndIf
		\If {$t$-th candidate point $\bm{p}_{i_t}$ doesn't exist}
		\State \textbf{break}
		\EndIf
		\State conflict = false
		\ForAll {neighbor cells $j$}
		\If {$j$ not a valid cell}
		\State \textbf{break}
		\EndIf
		\If {$j$ has no sample $\bm{s}_j$ yet}
		\State \textbf{break}
		\EndIf
		\If {dist($\bm{p}_{i_t}$, $\bm{s}_j$) < $d$}
		\State conflict = true
		\EndIf
		\EndFor
		\If {conflict == false}
		\State sample for cell $i$:  $\bm{s}_i = \bm{p}_{i_t}$
		\State put $\bm{s}_i$ in $\mathbb{S}$
		\EndIf
		\EndFor
		\EndFor
	\end{algorithmic}
	\caption{Uniform surface sampling}
	\label{alg:surface}
\end{algorithm}

The main sampling algorithm consists of several trials $n$, which is another controllable parameter. A higher number of trials $n$ makes it more likely to find a sample for each valid cell but needs more computation time. We employ a default value of $n = 10$. In each trial $t = 1,\ldots ,n$, each valid cell is processed. If the cell already contains a valid sample it is skipped. If there are no more candidate positions in $\mathbb{P}$, the cell can also be neglected. Otherwise, neighboring cells are checked to see if any already sampled position is too close to the current candidate position $\bm{p}_{i_t}$. If this is not the case $\bm{p}_{i_t}$ is added to the set of samples $\mathbb{S}$. The whole sampling algorithm is shown in Algorithm \ref{alg:surface}. For a parallel version and further details, we refer the reader to \cite{Bowers10}.

\subsection{Volume Sampling}
\label{subsec:volume}
In contrast to surface sampling, volume sampling is used to define initial positions of simulation particles, whether it may be free-moving particles like fluid, gas, or granular material, or a rigid-body representation in a unified solver like unified particle physics \cite{UPP14}. Therefore the requirements for the sampling are different. For initial positions the particles mustn't be primarily in an invalid state, meaning that the particles are not allowed to overlap. Overlapping initial positions can lead to high correction forces/position changes, causing the simulation to explode.

Particles that represent a rigid-body should maintain a constant distance from each other during simulation. Therefore it makes sense to sample them closely and uniformly inside the volume $V$. The goal is to find a set of sample positions $\mathbb{S}$, that is spaced out on an equidistant grid with a minimal side length of $2 \cdot r$, where $r$ is the particle radius. For deciding which positions are inside the Volume $V$ we use a signed distance field (SDF). It is defined as a signed distance function $\Psi \; : \; \mathbb{R}^3 \rightarrow \mathbb{R}$, 
\begin{align*}
\Psi(\bm{p}) &= \text{sgn}(\bm{p}) \inf_{\bm{p}^{*} \in \partial V}  \left\Vert \bm{p} - \bm{p}^{*}\right\Vert \quad \text{with}\\
\text{sgn}(\bm{p}) &= \begin{cases}
-1 \quad &\bm{p} \in V\\
1 \quad &otherwise
\end{cases}
\end{align*}
measuring the Euclidean distance of a point $\bm{p} \in \mathbb{R}^3$ to the nearest point $\bm{p}^{*}$ on the object surface $\partial V$. Generating such a parametric SDF representation of an arbitrary closed Volume $V$ is a topic on its own and beyond the scope of this paper. In \textsc{Leaven} we use a grid-based SDF with hierarchical cell size and polynomial degree refinement based on polynomial fitting as described in \cite{Discregrid2017}. The general idea behind this is to fit polynomial functions in cells of axis-aligned hexahedral grids defined by the input meshes. The algorithm iteratively refines the SDF either by spatial subdivision or cell-by-cell adjustment of the degree of the polynomial.

\begin{figure}
	\centering
	\includegraphics[width=1.0\linewidth]{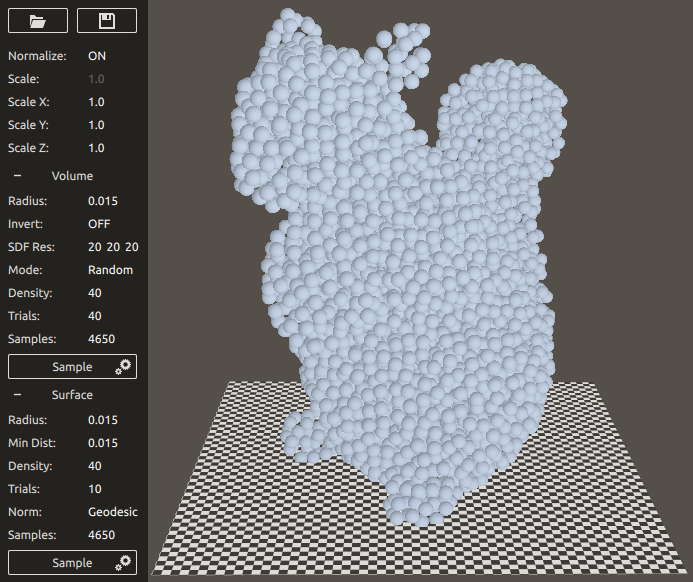}
	\caption{\textsc{Leaven}'s application window with a triangle mesh sampled with 4650 randomly distributed particles through the volume.}
	\label{fg:ui}
\end{figure}

After generating the SDF, the bounding box containing $V$ is divided into grid cells similar as described in Section \ref{subsec:surface} but with a cell size of $2 \cdot r$. For each cell $i$ one sampling particle is added to the cell center at position $\bm{p}_i$. Then the signed distance function is evaluated as position $\bm{p}_i$. When $\Psi(\bm{p}_i) < 0$ the position is inside the Volume $V$ and therefore it is a valid sampling position $\bm{s}_i$, that can be added to the sampling set $\mathbb{S}$ (see Algorithm \ref{alg:volume}).

\begin{figure*}
	\centering
	\includegraphics[width=1.0\linewidth]{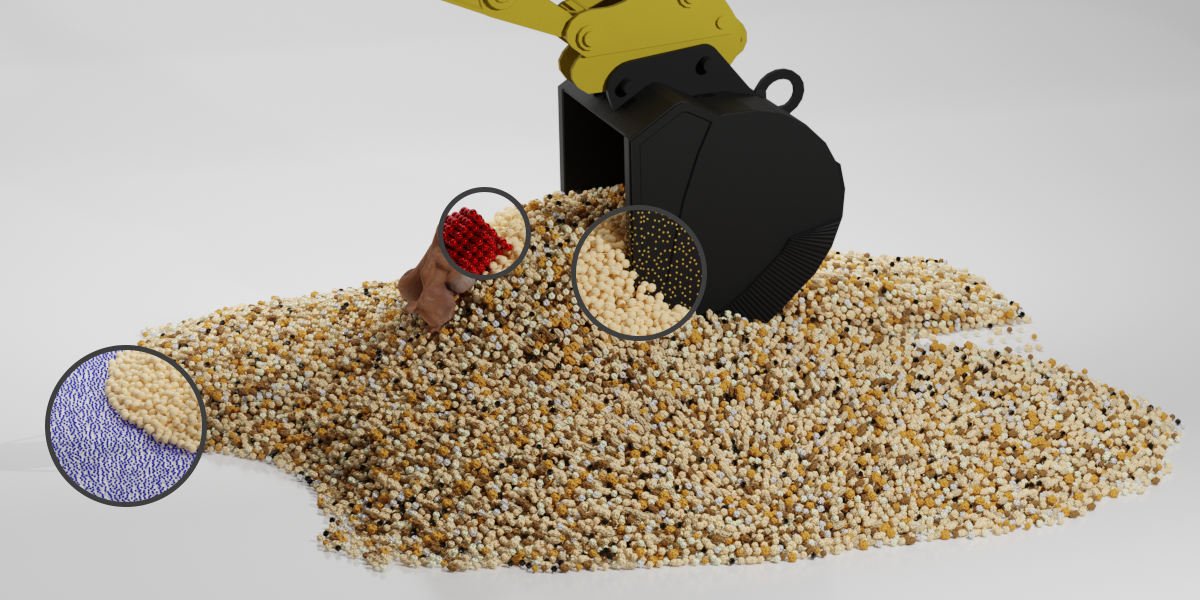}
	\caption{Volume and surface samplings used in a unified simulation framework simulating a granular material and a rigid-body (red) moved by an excavator.}
	\label{fg:usage}
\end{figure*}

\begin{algorithm}
	\centering
	\begin{algorithmic}[1]
		\State generateSDF()
		\ForAll {cells $i$}
		\State position $\bm{p}_i$ = cellCenter
		\If {$\Psi(\bm{p}_i)$ < 0}
		\State sample $\bm{s}_i$ = $\bm{p}_i$
		\State put $\bm{s}_i$ in $\mathbb{S}$
		\EndIf
		\EndFor
	\end{algorithmic}
	\caption{Grid-based volume sampling}
	\label{alg:volume}
\end{algorithm}  

Particles that represent parts of fluids, gases, or granular materials, in general, change their distance to each other during the simulation. Sampling them in regular patterns, like previously, can introduce disturbing visual artifacts at the beginning of a simulation. Therefore we modify the concept of Bowers et. al. for surface sampling \cite{Bowers10} explained in Section \ref{subsec:surface} to present a novel approach for generating random samples inside a 3D volume that satisfies blue-noise criteria.

Our modified sampling method divides the bounding box of $V$ into uniform grid cells with a side length of $\frac{2 \cdot r}{\sqrt{3}}$ resulting in $N$ cells. We fill the set of candidate positions $\mathbb{P}$ by picking uniformly distributed positions within the whole bounding box and only adding the positions $\bm{p}$ to $\mathbb{P}$ that lie inside the volume (i.e. $\Psi(\bm{p}_i) < 0$). In this way, we generate $\rho_V \cdot N$ samples, where the number of samples generated is proportional to the number of grid cells in the bounding box. The proportionality constant $\rho_V$ is a controllable density parameter. We recommend this parameter to be equal to the number of trials $n$. The rest of the sampling process is the same as for the surface sampling algorithm with Euclidean distance norm (see Algorithm \ref{alg:surface}).

\section{Application}
\label{sec:app}

As mentioned before, the main goal with \textsc{Leaven} is to provide an easy-to-use tool suitable for beginners and advanced users in particle-based simulation. In the following, we will briefly explain the settings the user can manipulate in the application\footnote{https://github.com/a1ex90/Leaven}.

Figure \ref{fg:ui} shows the \textsc{Leaven} user interface (UI). On the righthand side of the UI is the viewer. It displays the result of the sampling process. The depicted 3D model can be manipulated by arcball rotation, zooming, and panning. The lefthand side of the UI contains the settings for the sampling process. When mesh normalization is turned on, the mesh is uniformly scaled to fit inside a bounding box with a side length of 1. The mesh can be scaled further up or down by a factor. It is also possible to scale each axis individually, for example, to easily generate rectangular boxes from a unit cube for non-cubic simulation domains.

For volume sampling, the radius of the sampling particles can be specified. When sampling is inverted, the volume between the bounding box and the mesh outside is sampled. The SDF Resolution defines the accuracy of the parametric volume representation. For objects with finer details, a higher resolution is needed. The grid mode samples the particles uniformly and aligned on a grid inside the volume. Random mode generates randomized sampling inside the volume that satisfies blue noise criteria as described in Section \ref{subsec:volume}. In random mode the controllable density parameter $\rho_V$ and the number of trials $n$ can be set.

For surface sampling, the radius of the displayed sampling particles can also be specified. This only affects the visual appearance of the particles in the application's preview display and should not be confused with the \emph{Min Dist} parameter, which controls the minimum distance $d$ between particles that influences the sampling (see right side of Figure \ref{fg:ui}). For boundary handling, a good rule of thumb for choosing $d$ is to use the same size as the simulation particle radius $r$. Furthermore, the controllable density parameter $\rho_S$ and the number of trials $n$ can be chosen (see Section \ref{subsec:surface}). It is also possible to switch between Geodesic or Euclidean distance norm for sampling particle distances. For more fine details in the mesh, the geodesic norm should be the one in favor.


\section{Usage Example}
\label{sec:usage}

Figure \ref{fg:usage} shows a simulation scene where our sampling has been applied to many various objects. The scene has been simulated with the unified particle physics \cite{UPP14} variant of position-based dynamics \cite{PBD07}. A granular material with friction, behaving like sand, is simulated. Initial particle positions (sand-colored) are sampled with \textsc{Leaven}'s volume sampling random mode on. The rigid-body, a squirrel, is simulated within the same framework with a two-way coupling with the granular material. It is sampled with grid-based volume sampling (red). The excavator (yellow dots), as well as the domain boundaries (blue dots), are surface sampled with a minimal distance between samples equal to the particle radius.


\section{Conclusion}
\label{sec:conclusion}
With this paper we contribute an easy-to-use application for representing surfaces and volumes by a set of particles. We used a state-of-the-art surface sampling algorithm \cite{Bowers10} and present a novel approach for generating volume samples that satisfy blue noise criteria by using an SDF representation of the volume. Furthermore, we give a beginner-friendly introduction to the field of sampling techniques tailored for the needs of particle-based simulation.


\bibliographystyle{plain}

\footnotesize
\bibliography{references}

\end{document}